\documentclass[preprint,showpacs,preprintnumbers,amsmath,amssymb]{revtex4}
\usepackage{amsmath,amssymb,graphicx} 
\usepackage {amssymb}
\newcommand{\nc}{\newcommand}
\nc{\ba}{\begin{eqnarray}}
\nc{\ea}{\end{eqnarray}}
\newcommand\bb{{\mathbf{b}}}
\newcommand\A{{\mathbf{a}}}

\newcommand\s{\sigma}

\nc{\ga}{\gamma}
\nc{\x}{{\bf x }}
\nc{\kk}{{\bf k }}
\nc{\f}{{\bf f }}
\nc{\e}{{\bf e }}
\nc{\T}{ \theta (s_i (t)- \s) }
\nc{\TT}{ \theta (s_i (t_{ r \, i } )- \s) }
\nc{\br}{   (s_i (t)- \s)  }
\nc{\Kp}{ \hat{K}_+  }
\nc{\Km}{ \hat{K}_-  }
\nc{\Kpj}{ \hat{K}_{ + }^{j}  }
\nc{\Kmj}{ \hat{K}_{ - } ^{j}}
\nc{\cA}{{ \cal A} }
\nc{\cB}{{ \cal B} }

\input{epsf.sty}

\begin{document}

\title{Gravitational Radiation by Cosmic Strings in a Junction }

\author{R. Brandenberger}  \email{rhb@physics.mcgill.ca}
\affiliation{ Physics Department, McGill University,   Montreal, H3A 2T8, 
Canada., and
Institute of High Energy Physics, Chinese Academy of Sciences, P.O. Box 918-4, 
Beijing 100049, P.R. China}

\author{ H. Firouzjahi} \email{firouz@ipm.ir}
\affiliation{ School of Physics, Institute for Research in Fundamental Sciences (IPM), Tehran, Iran.}

\author{J. Karouby }\email{karoubyj@physics.mcgill.ca}
\affiliation{\,  Physics Department, McGill University,   Montreal, H3A 2T8, 
Canada. }

\author{S. Khosravi} \email{khosravi@ipm.ir}
\affiliation{ Physics Department, Faculty of Science, Tarbiat Mo'alem University, Tehran, Iran, and
School of Astronomy, Institute for Research in Fundamental Sciences(IPM), Tehran, Iran.
}

\vspace{2cm}

\begin{abstract}
\vspace{0.3cm}
The formalism for computing the gravitational power radiation from excitations on cosmic 
strings forming a junction is presented and applied to the simple case of co-planar strings 
at a junction when the excitations are generated along one string leg. 
The effects of polarization of the excitations and of the back-reaction of the
gravitational radiation on the small scale structure of the strings are studied.


\vspace{0.3cm}

\end{abstract}

\maketitle

\section{Introduction}

In models of brane inflation cosmic strings are produced (for a review see \cite{HenryTye:2006uv}). 
This has led to a revival of interest in cosmic strings (see e.g. \cite{Kibble:2004hq,ACD,Mairi}). 
Cosmic strings forming in the context of brane models can take the form of Fundamental strings (F-strings), D1-branes (D-strings) or their bound states, ((p,q) strings). A (p,q) string is a bound state 
of $p$ F-strings and $q$ D-strings. Networks of stringy cosmic strings which can involve
strings with different values of $p$ and $q$ have features unlike those of simple gauge theory
strings. Unlike U(1) gauge theory cosmic strings which inter-commute when they intersect, in
the case of cosmic (p,q) strings there are conservation laws which prevent the inter-commutation
of strings with different values of $p$ and $q$. Instead, a string junction can be formed. For
example, a $p$ string and a $q$ string can join at a junction to form a (p,q) string. 
The construction of cosmic strings with junctions and its cosmological implications were studied
in  \cite{Copeland:2006eh, Copeland:2006if,Shlaer:2005ry, Copeland:2007nv, Brandenberger:2007ae, Urrestilla:2007yw, Leblond:2007tf, 
Dasgupta:2007ds, Suyama:2008ty, Davis:2008kg,  Rajantie:2007hp,  Avgoustidis:2007aa}.

Gravitational wave (GW) emission from loops and cusps of cosmic strings has been
studied (for a comprehensive review see \cite{Hindmarsh:1994re, book}
and for more recent analyses see \cite{Damour:2004kw,Xiemens}). A straight infinite 
string does not emit GW. This is because to emit GW, as we shall 
explicitly see in the next section, both left-movers and right-movers should be present 
on the string world sheet. In a network of cosmic strings, it is quite natural to expect that 
wiggles of different wavelengths are generated on the world sheet of an infinite string. 
These wiggles, for example, are left over from times when the correlation length of
the string network was much smaller, or are remnants of string inter-commutations which
took place in the past. These wiggles cause the GW emission from long strings and can 
smooth out the wiggles of the string world sheet. GW emission from wiggles on a straight
string were studied in \cite{Sakellariadou:1990ne, Hindmarsh:1990xi, OS}. In particular, in 
\cite{OS} left-moving and right-moving wave-trains of different wavelengths and amplitudes 
on an infinite string were considered. It was shown that when the wavelengths and the amplitudes 
of the wave-trains are comparable, the GW emission is mainly from lower harmonics and is 
proportional to the frequency of the wave-trains. This indicates that excitations of higher 
frequency die out faster than excitations of shorter frequencies. On the other hand, when 
the wavelengths and amplitudes of the wave-trains are much different, then GW emission is exponentially suppressed. 

As mentioned above, the formation of junctions is a generic feature of networks of cosmic superstrings. With this motivation, in this paper, we consider gravitational radiation from strings at a junction. 
As we shall see, the presence of the junction leads to mixing of left and right-moving
excitations on the string which is the necessary criterium for the emission of GW. In 
Section 2, we present the setup of our study. In Section 3 we study three examples. The 
first example is GW emission from a semi-infinite string attached to a rigid wall. The second 
example corresponds to GW emission from a stationary junction. The third example concerns 
GW emission from a non-stationary junction. As we shall see, the expressions for the
gravitational wave power radiated has a similar form in all three examples. We discuss
our results and summarize our conclusions in Section 4.  

\section{The Setup}

Our setup consists of semi-infinite strings forming a stationary junction. The formalism in this 
section is valid for any number of semi-infinite strings meeting at a junction. However,
to be specific, in our study we shall focus on  the simple example where three semi-infinite 
strings form a stationary junction.  

The world-sheet of each string is described by a temporal coordinate $ \tau $ and
a string length parameter $\s$. The induced metric $\ga_{i\, ab}$ on each string is given by 
\ba
\label{gamma}
\ga_{i\, ab} = g_{\mu \nu}  \,  \partial_{a } \,  X_{i}^{\mu} \partial_{b} \, X_{i}^{\nu}  \, .
\ea
Here and in the following, we reserve $\{a, b\}= \{ \tau, \sigma \}$ for the string world-sheet indices 
while Greek indices represent the four-dimensional space-time coordinates. 
Furthermore, $X_{i}^{\mu}$ stands for the position of the $i$-th string in four space-time
dimensions.  

We impose the conformal temporal gauge on the string world-sheet for which 
$X_{i}^{0}= t=\tau$ and $\ga_{ i\, 0\s}=0$. This is equivalent to
\ba
\label{gauge1}
\dot \x_{i}  \, . \,  \x_{i}' =0 \quad , \quad \dot \x_{i}^{2} + \x_{i}'^{2} =1 \, .
\ea
Here an overdot and a prime denote derivatives with respect to $t$ and $\s$, respectively, 
while  $\x_{i}$ represent the spatial components of the $i$-th string.

For the components of the induced metric on the string  world sheet we obtain
\ba
\ga_{ i\, 00} = 1- \dot \x_{i}^{2}     \quad , \quad \ga_{i\, \s \s} = - \x_i'^{2} = -\ga_{i\, 00} \, .
\ea

We start with the following action
\ba
\label{action1}
S=-\sum_{i} \mu_{i} \int d\, t \, d\, \s \sqrt{-|\ga_{i} |}  \,  \T \, 
\ea
where $|\gamma_i|$ is the determinant of the world sheet metric of the $i$-th string.
We are using the convention that the position of the junction on the $i$-th string is given by 
$s_{i}(t)$. It is assumed that $\sigma$ is increasing towards the junction. We can impose a 
lower cutoff on $\sigma$, which  would correspond to the physical length of the string 
under consideration. The equation of motion for $s_{i}(t)$ and the conditions for junction 
formation have been studied in \cite{Copeland:2006eh, Copeland:2006if, Copeland:2007nv}.

The energy-momentum tensor for the action given by Eq. (\ref{action1}) is obtained by 
varying the action with respect to the background metric $g_{\mu \nu}$, with the
result
\ba
\label{var1}
\delta_{g_{\mu\nu}} S= -\frac{1}{2} \sum_{i} \mu_{i} \int d\, t \, d\, \s \sqrt{-|\ga_{i} |}  \, 
\ga^{ab}   \partial_{a } \,  X_{i}^{\mu} \partial_{b} \, X_{i}^{\nu}  \, \T\,  \delta g_{\mu \nu}
\equiv  -\frac{1}{2} \int d^{4} x  \, T^{\mu \nu} \,  \delta g_{\mu \nu} \, 
\ea
which gives
\ba
\label{stress}
T^{\mu \nu} (x)&=& \sum_{i} \mu_{i} \int d\, t \, d\, \s \sqrt{-|\ga_{i} |}  \, 
\ga^{ab}    \,  X_{i, \, a}^{\mu}  \, X_{i,  \, b}^{\nu}  \,
 \T \, \delta^{(4)} (x - X_{i} ) \nonumber\\
 &=&
 \sum_{i} \mu_{i} \int d\, t \, d\, \s (\dot X^{\mu} \dot X^{\nu} - X'^{\mu} X'^{\nu}) 
 \,  \T \, \delta^{(4)} (x - X_{i} ) \, .
\ea

Having obtained the energy-momentum tensor, we can use the standard formalism for 
calculating GW emission from a source \cite{Weinberg} .The derivation
in \cite{Weinberg} is for a source which is localized in space. To justify the application of
the formalism to the case of a long string, we can imagine considering first short wave-trains
on the string, in which case the formalism of \cite{Weinberg} applies as
it was initially derived, and then taking the limit in which the length of the wave trains
increases. This limit does not lead to any problems when applying for formalism . 
According to this formalism, the power emitted in direction  $\kk$ per solid angle 
$\Omega$, integrating over the frequencies $\omega$ of the emitted waves,  
is given by
\ba
\label{power}
\frac{d E}{d \Omega} = 2 G \int_{0}^{\infty} d \omega \omega^{2} 
\left[ {T^{\lambda \nu}}^{*} (k) \,  T_{ {\lambda \nu} } (k) 
 - \frac{1}{2} | T^{\lambda}_{\lambda} (k) |^{2} \right] \, ,
\ea
where $G$ is Newton's gravitational constant and $T_{ {\lambda \nu} } (k) $ is the 
Fourier transform of $T_{ {\lambda \nu} } (t , \x) $
\ba
T_{\mu \nu}(k)=  \frac{1}{2 \pi} \int d^{4} x \,  T_{\mu \nu} (x) \,  e^{i k.x} \, .
\ea

In conformal temporal gauge the solution of the string equations of motion
\ba
{\ddot X}^{\mu} -{X''}^{\mu} =0 \,  
\ea
can be represented by the combination of left-moving and right-moving modes:
\ba \label{ab}
X^{\mu}_{i} =\frac{1}{2} \left( a_i^{\mu}(v) + b_i^{\mu}(u)   \right)   \quad , \quad
 {a'_i}^{2} = {b'_i}^{2}=0 \, .
\ea
where $v = \sigma+t$ and $u= \sigma-t$ are the light-cone coordinates.

Since we need the components of the energy-momentum tensor in Fourier space, it is 
useful to replace the $\theta$ function by its Fourier representation which is
\ba
\theta(x)= \frac{1}{2 \pi \, i} \int_{-\infty}^{\infty} \frac{ d  \ell \, e^{i \ell x}}{\ell - i \varepsilon}
\quad, \quad  \varepsilon \rightarrow 0^{+} \, .
\ea
Inserting this into  Eq. (\ref{stress}), we find
\ba
\label{Tmuk1s}
T^{\mu \nu} (k)= \sum_{j} \frac{\mu_{i}}{8 \pi} \frac{1}{ 2 \pi i} 
  \int_{-\infty}^{\infty} \frac{ d \ell }{\ell - i \varepsilon} &&\int du  dv 
   (a_{j}'^{\mu} b_{j}'^{\nu} + a_{j}'^{\nu} b_{j}'^{\mu}) \nonumber\\
&&\times \, \exp{  \left[   i \ell   s_{j} (u,v)-  \frac{i \ell}{2} (u + v)  
+ \frac{i}{2} k. (a_{j} + b_{j})  \right] } .
\ea

In the first two examples in the following section, we consider cases when the junction
remains stationary, corresponding to $s_{i}  =0$. In this case, one obtains
\ba
\label{Tmuk1}
T^{\mu \nu} (k)= \sum_{j} \frac{\mu_{j}}{8 \pi} \frac{1}{ 2 \pi i} \, 
 \int_{-\infty}^{\infty} \frac{ d \ell }{\ell - i \varepsilon}
 \left(  A^{\mu}_{j}(k, \ell) B^{\nu}_{j}( k, \ell)
+  A^{\nu}_{j}(k, \ell) B^{\mu}_{j}(k, \ell) \right)
\ea
where
\ba
\label{AB}
A^{\mu}_{j}(k, \ell) \equiv 
\int_{-L/2}^{L/2} dv \, a_{j}'^{\mu}(v)\,  \exp \left [{i  k. a_{j}(v)/2- i \ell v/2}  \right]
\nonumber\\
B^{\mu}_{j}( k, \ell) \equiv 
\int_{-L/2}^{L/2} du \, b_{j}'^{\mu}(u)\,  \exp \left[ {i  k. b_{j}(u)/2- i \ell u/2}  \right] \, ,
\ea
where $L$ is the physical length of the string being considered (since the GW
emission comes from regions where the wave trains are non-vanishing, effectively
$L$ can be taken as the length on the string which corresponds to the region where
the wave-trains are localized), and we assumed the mid point of the string is at
world sheet coordinates $u=v=0$.

To calculate $A_{i}^{\mu}(k, \ell)$ and $B_{i}^{\mu}(k, \ell)$ we follow the formalism of  \cite{OS}. 
We assume that on each string there are left-moving and right-moving wave-trains of
lengths  $L_{i}= N_{a}^{i} \lambda_{a}^{i} $ and $L_{i}= N_{b}^{i} \lambda_{b}^{i} $ for 
integers $N_{a}^{i}$ and $N_{b}^{i}$, where $\lambda_{a(b)}^{i}=  2\pi/ \kappa_{a(b)}^{i}$ are the wavelengths of the left(right)-moving wave-trains on each string.

We are interested in GW emission from the excitations of the strings and neglect the 
contributions of the straight parts of the strings to $T^{\mu \nu}$.  To establish our notation, 
the contributions to a quantity $Q$ from the string excitations are denoted by $\delta Q$.
For example, $a_{i}^{\mu} \rightarrow a^{\mu}_{i} + \delta a_{i}^{\mu}$ and so on. Discarding 
the contributions from the straight parts of the strings (which do not contribute to
gravitational radiation), one obtains for the fluctuating part of $T^{\mu \nu}$:
\ba
\label{Tmuk2}
\delta T^{\mu \nu} (k)= \sum_{j} \frac{\mu_{j}}{8 \pi} \frac{1}{ 2 \pi i} \, 
 \int_{-\infty}^{\infty} \frac{ d \ell }{\ell - i \varepsilon}
(\delta A_{j}^{\mu} \delta B_{j}^{\nu} + \delta A_{j}^{\nu} \delta B_{j}^{\mu} ) \, ,
\ea
where
\ba
\delta A_{j}^{\mu}(k , \ell) =  \int_{- L/2}^{L/2}  dv  \,  e^{iv \Kpj /2}
  \left(  \delta a_{j}'^{\mu} + \frac{i}{2} a_{j}'^{\mu} k. \delta a_{j}  \right)
\ea
and $  \delta B^{\mu}(k , \ell)  $ is given by a similar expression. Here
\ba
\Kpj \equiv K_{+}^{j} -\ell
 \quad , \quad 
K_{ +}^{j} \equiv k.a_{j}'  
 \ea
with $\Km$ and $K_{-}$ defined similarly for the right-movers.

For the case where $s_{i} \neq 0$, such as in the third example in the next section, 
we obtain
\ba
\label{Tmuk2s}
&&\delta T^{\mu \nu}(k)= \sum_{j} \frac{\mu_{j}}{8 \pi} \frac{1}{ 2 \pi i} \, 
 \int_{-\infty}^{\infty} \frac{ d \ell }{\ell - i \varepsilon} \left\{   \left. \right.
(\delta A_{j}^{\mu} \delta B_{j}^{\nu} + \delta A_{j}^{\nu} \delta B_{j}^{\mu} ) \right.\nonumber\\
&+& \left. i \ell \int du dv  
s_{j} e^{i \Kpj v/2}  e^{i \Kmj u/2} \left[
 ( a_{j}'^{\mu}  \delta b_{j}'^{\nu} + b_{j}'^{\nu}  \delta a_{j}'^{\mu}  + \mu \leftrightarrow \nu )
+ \frac{i}{2} k. (\delta a_{j} + \delta b_{j} )( a_{j}'^{\mu} b_{j}'^{\nu} +  a_{j}'^{\nu} b_{j}'^{\mu} ) \right]  \right.
\nonumber\\
&-& \frac{\ell^{2}}{2} \int du \, dv \, s_{j}^{2}   e^{i \Kpj v/2}  e^{i \Kmj u/2}  
( a_{j}'^{\mu} b_{j}'^{\nu} + a_{j}'^{\nu} b_{j}'^{\mu} )
\left.
\right\} 
\ea

\section{Examples}

In this section we employ the formalism presented in the last section to calculate GW emission 
for different examples. 

\subsection{A semi-infinite string attached to the wall}

The first example we would like to consider is gravitational radiation from a semi-infinite string 
attached to a rigid wall. An incoming perturbation is coming from infinity, hits the wall and 
gets reflected. This creates wave-trains of both left-mover and right-mover on the string. This 
problem is in spirit very similar to the problem of two left-moving and right-moving wave-trains propagating on an infinite string studied by Siemens and Olum \cite{OS}. 

The non-fluctuating string configuration is given by
\ba
a'^{\mu}= (1, \e)  \quad , \quad b'^{\mu}= (-1, \e) 
\ea
where the unit vector $\e $ represents the orientation of the string. We could simply take the 
vector $\e$ to be along the $z$ axis. However, in order to establish a formalism which can 
also be applied to the next examples involving strings oriented in different directions, we 
keep the vector $\e$ unspecified. The perturbations on the string are given by
\ba
\delta a'^{\mu} =  \epsilon_{a} \f \, \cos (\kappa_{a } v)  \quad , \quad 
\delta b'^{\mu} = \epsilon_{b} \f \, \cos (\kappa_{a } u) 
\ea
where $\epsilon_{a(b)}$ are small numbers controlling the amplitude of the perturbations, 
$\kappa_{a(b)}$ are the frequencies of the left(right)-moving perturbations and $\f$ is a unit 
vector indicating the polarization of the perturbations with $\e . \f =0 $.
In this example  we know that  $\epsilon_{a}= \epsilon_{b}$ and $\kappa_{a}=\kappa_{b}$ . 
However, in order to keep the formalism general we have not made these identifications.

Calculating $\delta A^{\mu}$, one obtains
\ba
\label{deltaA1}
\delta {\vec A}= \delta A^{0} \left( \e - \frac{\Kp}{\kk .  \f}  \f       \right)
\quad , \quad
\delta A^{0}= 
 -4\,(-1)^{N_{a}}  \epsilon_{a} \, \kk . \f \, 
\, \frac{     \sin(L \Kp/4)        }{  \Kp^{2} - 4 \kappa_{a}^{2}}
\ea
with a similar expressions for $\delta B^{\mu}$ with $\Kp$ replaced by $\Km$.

To calculate $\delta T^{\mu \nu}(k)$, we need to plug $\delta A^{\mu}(k, \ell)$ and
$\delta B^{\mu}(k, \ell)$  into Eq. (\ref{Tmuk2}) and integrate over $\ell$. There are five
poles at
\ba
\ell_{1}= i \varepsilon \quad , \quad 
\ell_{2,3} = K_{+} \pm 2 \kappa_{a}  \quad , \quad 
\ell_{4,5} = K_{-} \pm 2 \kappa_{a} \, .
\ea
One can easily check that only the residue at $\ell_{1}$ gives a non-zero contribution
to $\delta T^{\mu \nu}(k)$ and the other residues vanish. For example, calculating the residue
at $\ell=\ell_{2}$, one obtains that  $  \delta T^{\mu \nu}(k) \propto \sin (L \kappa_{a}/2  )= \sin(N_{a} \pi)=0 $. 

Calculating the residue at $\ell=\ell_{1}$, one obtains
\ba
\label{deltaT1}
\delta T^{\mu \nu} (k) = \frac{\mu_{1}}{8 \pi} 
\left[\delta A^{\mu}(k, 0) \delta B^{\nu}(k,0) + \delta A^{\nu}(k,0) \delta B^{\mu}(k,0) \right] \, .
\ea
Noting that  $\delta A^{\mu}(k, 0)$ and $\delta B^{\mu}(k, 0)$ are real, one obtains
\ba
\label{dE}
\frac{ d E}{ d\Omega \, d \omega} &=& \frac{ G \mu_{1}^{2}  }{ 16 \pi^{2} } \omega^{2} 
\delta A(k,0)^{2}  \delta B(k,0)^{2}  \nonumber\\
&=&\frac{ 16\, G \mu_{1}^{2}  }{ \pi^{2} } \omega^{2} 
\epsilon_{a}^{2} \epsilon_{b}^{2} \, K_{+}^{2} \,  K_{-}^{2}   \, 
\frac{     \sin^{2}(L K_{+}/4)        }{ ( K_{+}^{2} - 4 \kappa_{a}^{2})^{2} }  \, 
\frac{     \sin^{2}(L K_{-}/4)    }{ ( K_{-}^{2} - 4 \kappa_{b}^{2})^{2} }
\ea
One interesting result which emerges from the above is that the combination $\kk.\f$ drops 
from the numerator and denominator of the above expression. This indicates that the GW 
power emission is independent of the polarization of the incoming waves. This feature will also 
show up in next example.

To exploit the symmetry of the problem, now we assume that the string is oriented along the $z$ 
axis, so $\kk . \e = \omega \cos \theta$, where $\theta$ is defined as the angle between the 
vector $\kk$ and the orientation of string. On the other hand, calculating $K_{+}$ and $K_{-}$, 
we get
\ba
\label{Kpm}
K_{+} = \omega (1- \cos \theta) \quad , \quad
K_{-} = -\omega (1+ \cos \theta) \, .
\ea

As in \cite{OS}, changing the coordinates from $(\omega, \cos \theta)$ to $(K_{+}, K_{-})$, and 
noting that $2 \omega = K_{+}- K_{-}$ and $d \omega  \, d(\cos \theta) = d K_{+} d K_{-}/2 \omega$, 
one obtains
\ba
\label{dEdphi0}
\frac{ d E}{ d\phi }  = \frac{ 4\, G \mu_{1}^{2}  }{ \pi^{2} }  \epsilon_{a}^{2} \epsilon_{b}^{2} 
\int dK_{+} d K_{-}  (K_{+}- K_{-})  K_{+}^{2} K_{-}^{2} \,  
\frac{     \sin^{2}(L K_{+}/4)        }{ ( K_{+}^{2} - 4 \kappa_{a}^{2})^{2} }  \, 
\frac{     \sin^{2}(L K_{-}/4)    }{ ( K_{-}^{2} - 4 \kappa_{b}^{2})^{2} } \, .
\ea
Here $\phi$ is defined as the azimuthal angle around the string.

The above integral can be performed using  the following approximations for large $N_{a}$ \cite{OS}
\ba
\frac{\sin^2{(N_{a} x^{a}/2)}}{ \sin^2{(x^{a}/2)}} = 2 \kappa_{a} N_{a} \sum_{n=-\infty}^{\infty}
\delta (K_{+} - 2 n \kappa_{a} ) \, 
\ea
where $x_{a} \equiv  \lambda_{a} K_{+}/2$.  A similar identity also holds for right-movers
with $K_{+} \rightarrow K_{-}, N_{a} \rightarrow N_{b}$ and $n\rightarrow m$.
Using this identity, Eq. (\ref{dEdphi0}) yields
\ba
\label{dEdphi}
\frac{ d E}{ d\phi }  = \frac{ 2 G \mu_{1}^{2}}{ \pi^{2}} \frac{N_{a} N_{b}}{ \kappa_{a} \kappa_{b}}
\sum_{m,n} \frac{  n^{2} m^{2} (n \kappa_{a} - m \kappa_{b} ) \sin^{2}(n \pi) \sin^{2} (m \pi)  }{ (n^{2}-1)^{2} (m^{2}-1)^{2}    } \, .
\ea

Knowing that $K_{+}\ge0, K_{-}\le0$, one can see that only $n=-m=1$ contribute in the 
summation in Eq. (\ref{dEdphi}) and one obtains
\ba
\label{dE1}
\frac{ d E}{ d\phi }  = \frac{G \mu_{1}^{2}}{8} N_{a} N_{b } \pi^{2} \, 
\epsilon_{a}^{2}  \epsilon_{b}^{2} 
  \frac{\kappa_{a} + \kappa_{b}}{ \kappa_{a} \kappa_{b}} \, .
\ea
We are interested in the power radiated per unit of length, $d P/d l$, which is obtained by 
dividing the above expression by the world-sheet volume of the string, $L_{a} L_{b}/2$. 
After integrating over the angle $\phi$ and noting that $\epsilon_{a}= \epsilon_{b}$ and 
$\kappa_{a}=\kappa_{b}$, we obtain
\ba
\label{dPdz1}
\frac{ d P}{ d\, l }  = \frac{G \mu_{1}^{2} \pi}{16} \epsilon_{a}^{4}  \kappa_{a} \, .
\ea

Note that our result is identical to the power radiated per unit of length 
obtained in \cite{OS} for an infinite string. This is not surprising since
our semi-infinite string locally looks identical to an infinite string. The only
difference is at the junction with the wall - but since that point is not moving
it does not contribute to the power of gravitational radiation. Thus, we expect
that our result for a semi-infinite string agrees with that of \cite{OS} for an infinite 
string (there is a factor of 1/2 mistake in the original version of Eq. (72) of 
\cite{OS} which is corrected in the printed version. Taking that into account, our result
here agrees with their Eq. (36)).

As in \cite{OS} the power radiation is dominated by lower harmonics. Also for $n=-m=1$ 
we note that $\omega = 2 \kappa_{a}$, so the frequency of the radiation is  twice of the 
frequency of the incoming wave.


\subsection{Strings at a junction}

In this section we consider the problem of GW emission from strings at a junction. 
Three semi infinite strings form a stationary junction. There is an incoming right moving 
excitation on one string, say String 1. After the wave hits the junction, part of it is transferred 
to Strings 2 and 3, while part of the incoming wave is reflected along String 1 (see 
\cite{Copeland:2006if} for the details of the dynamics).
Depending on the polarization of the incoming wave, the junction may stay stationary, 
corresponding to $\delta s_{i}(t)=0$, or the junction may dislocate along the strings 
corresponding to $\delta s_{i}(t)\neq 0$. 

To be specific, suppose that the strings are in the $x-y$ plane and their orientations are 
given by $\e_{i}=(\cos \theta_{i}, \sin \theta_{i} , 0)$ where $\theta_{i}$ is  the angle of 
the $i$-th string with the $x$-axis. This gives
\ba
\label{ab1}
a'^{\mu}_{i}  &\equiv&   (1, \A_{i} ) = (1,\e_{i}) \nonumber\\
b'^{\mu}_{i} &  \equiv&   (-1, \bb_{i} ) =  (-1,\e_{i}) \, ,
\ea
where $\A_{i}(\bb_{i})$ indicates the spatial part of $a_{i}^{\mu}(b_{i}^{\mu})$.
The relations $\sum \mu_{i} \cos \theta_{i} = \sum \mu_{i} \sin \theta_{i} =0 $ also must be
satisfied if the junction is to be stationary (this is due to the force balance condition).

Now suppose there is a small incoming excitation on one string, say String 1, with
\ba
\label{deltab1}
\delta \bb_{1}'(u)= \epsilon \,  \f_{1} \cos( \kappa u) \, , 
\ea
and $\delta \bb_{2}'= \delta \bb_{3}'=0$. Here $\epsilon\ll1 $ is a dimensionless parameter
controlling the amplitude of the perturbation.

\subsubsection{The case $s_{i}(t)=0 $}

The simplest case is when $\f_{1}=(0,0,1)$. This has the advantage that the junction does 
not dislocate on the strings: $ \delta \dot s_{i}=0$ \cite{Copeland:2006if} and we can use 
the formalism developed in the previous section.  Following \cite{Copeland:2006if}, one finds 
\ba
\label{deltaa}
\delta \A_{1}' = \frac{\nu_{1}}{\mu} \epsilon  \, \f_{1} \cos (\kappa v) \quad, \quad
\delta \A_{2}' = \delta \A_{3}'  = 
\frac{2\mu_{1}}{\mu} \epsilon \,  \f_{1} \cos (\kappa v) \, ,
\ea
where
\ba
\label{numu}
\mu \equiv \mu_{1}+ \mu_{2} +\mu_{3} \quad , \quad 
\nu_{1}\equiv \mu_{2} + \mu_{3} - \mu_{1} \, .
\ea

With this initial condition, one sees that only $\delta B^{\mu}_{1}$ is non-zero. Furthermore, 
$\delta T^{\mu \nu}(k)$ is as given in Eq. (\ref{deltaT1}), with $\delta A^{\mu}_{1}$ given as in
 Eq.(\ref{deltaA1}). Thus, $d E/d \Omega \, d\omega$ has the same form as 
 Eq. (\ref{dE}). We note that String 1 is in $x-y$ plane. But we can label the coordinate
 (or perform a coordinate transformation) such that String 1 is along the $z$- direction. 
To calculate the power radiated, we can simply use Eq. (\ref{dE1}) with
 the identification  $\epsilon_{b}= \epsilon$ and $\epsilon_{a}= \epsilon \nu_{1}/\mu$, and 
 as before, we  note that $\phi$ is defined as the azimuthal angle around String 1.
The power radiated per unit of length, using Eq. (\ref{dE1}),  therefore is
\ba
\label{dPdz2}
\frac{ d P}{ d\, l }  = \frac{G \mu_{1}^{2} \pi}{16}  \frac{\nu_{1}^{2}}{ \mu^{2}   }
\epsilon^{4}  \kappa \, .
\ea

One may wonder why this result has the same form as that in the previous example where 
a semi-infinite strings was attached to a rigid wall. The reason is that here the junction 
plays the role of the rigid wall. Indeed, the fact that the junction remains stationary makes 
this analogy more manifest. The effect of Strings 2 and 3 is to let parts of incoming waves 
be transferred to them. This has the effect that $\epsilon_{a} \neq \epsilon_{b}$.


\subsubsection{The case $s_{i}(t) \neq 0$}

Now we consider the case when the polarization of  the incoming perturbation is in the
$xy$-plane. Then $s_{i}$ will also oscillate and the junction does not stay at a fixed 
position on each string \cite{Copeland:2006if}. As in  \cite{Copeland:2006if} we assume
\ba
\f_{i} =(-\sin \theta_{i}, \cos \theta_{i} , 0)
\ea
and $\e_{i} . \f_{i} =0$ for each string. As before, the incoming perturbation is along String 1 
and is given by 
\ba
\delta \bb_{1}'(u)= \epsilon \f_{1} \cos( \kappa u) \, . 
\ea
Following \cite{Copeland:2006if} , one obtains
\ba
\label{deltaa2}
\delta \A_{1}' = -\frac{\nu_{1}}{\mu} \epsilon  \, \f_{1} \cos (\kappa v) \quad, \quad
\delta \A_{2,3}' = 
\frac{2\mu_{1}}{\mu} \epsilon \,  \f_{2,3} \cos (\kappa v) \, ,
\ea
and
\ba
\label{deltas}
\delta s_{1} =  \frac{(\mu_{2}-\mu_{3}) \nu_{1}}{\kappa \, \Delta } \epsilon \sin (\kappa t)
\quad , \quad 
\delta s_{2} = -  \delta s_{3} =
  -\frac{\mu_{1} \nu_{1}}{\kappa \, \Delta } \epsilon \sin (\kappa t)
\ea
with $\nu_{1}$ as given in Eq. (\ref{numu}) and $\Delta = \sqrt{ \mu \nu_{1} \nu_{2} \nu_{3}  }$, 
where $\nu_{2,3}$ are defined like $\nu_{1}$ with the appropriate permutations.

With these forms of $\delta s_{i}$, one can show (see the {\bf Appendix}) that both integrals 
containing the linear and the quadratic powers of $s_{i}(t)$ in $\delta T_{\mu \nu}(k)$ in 
Eq. (\ref{Tmuk2s}) vanish. 

One sees that $\delta T^{\mu \nu}$ is of the same form as Eq. (\ref{deltaT1}) 
with $\delta A^{\mu}_{1}$ given as in Eq.(\ref{deltaA1}) . Like in previous example, the effect
of the junction is to make $\epsilon_{a}\neq \epsilon_{b}$. On the other hand, since the amplitude
of $\delta a_{1}^{\mu}$ is the same as in the previous example where the polarization was along the
$z$ axis, we find that the power radiated is the same as before, given by Eq. (\ref{dPdz2}).

The fact that the power of gravitational radiation when the polarization $\f_{1}$ is coplanar with 
the strings  is the same as when the polarization is perpendicular to the plane of the strings may 
seem surprising. However, as mentioned in \cite{Copeland:2006if}, one can consider the 
excitations on the strings as the propagation of massless particles. Using
conservation of energy, one can check that the transmission and reflection indices for
both polarizations are the same.


\section{Discussion}

In this paper, gravitational wave (GW) emission from strings at a stationary junction has
been studied. We considered the simple case when three co-planar semi-infinite strings
form a stationary junction. A purely left-moving wave, excited on one string, travels towards
the junction. Part of it is reflected from the junction while the rest is transferred to other strings.
The role of the junction therefore is to mix the left-moving and right-moving excitations 
which are necessary for GW emission.  

We found that power of gravitational radiation is independent of the polarization of the 
incoming wave. Furthermore, its magnitude is proportional to the frequency of the 
incoming wave. This means that excitations of higher frequencies (shorter wavelengths) die 
out faster than excitations with lower frequencies (longer wavelengths).

In \cite{OS, Siemens:2002dj, Polchinski:2007rg} the gravitational back-reaction effects 
on the small  scale structure present on a long string are studied. Here we shall briefly 
apply their formalism to our case. An excitation of 
the form Eq. (\ref{deltab1}) leads to a change $\delta \mu$ in the mass per unit
length of the string. This change takes the form
\ba
\delta \mu \sim \mu_{1} \epsilon^{2} \, .
\ea
The energy loss via gravitational radiation given by Eq. (\ref{dPdz1}) or  Eq. (\ref{dPdz2}) 
leads to a decrease of this contribution:
\ba
\frac{d}{d t} (\delta \mu) =- \frac{d P}{dz} =
- \frac{G \mu_{1}}{8 \lambda} \pi^{2} \frac{\nu_{1}^{2}}{\mu^{2}} \epsilon^{2} \delta \mu \, .
\ea
This differential equation has the solution
\ba
\delta \mu  \sim \exp{(-t/\tau)} \, ,
\ea
where
\ba
\tau= \frac{8 \lambda }{\pi^{2} G \mu_{1}} \frac{\mu^{2}}{\epsilon^{2} \nu_{1}^{2}}\, .
\ea
Excitations which survive until the present time $t_{0}$ are characterized by
$\tau > t_{0}$.  Taking $\epsilon \lesssim 1$,
the minimum wavelength of excitations that can survive is thus
approximately given by
\ba
\lambda_{min} \sim G \mu_{1} \frac{\nu_{1}^{2}}{\mu^{2}} \,  t_{0} \, ,
\ea
while on smaller scales the wiggles are exponentially suppressed.

In this analysis we have considered monochromatic wave in the form of Eq. (\ref{deltab1}).
In \cite{Siemens:2002dj} (see also \cite{Polchinski:2007rg}) the estimation of back-reaction 
was generalized to the case when higher harmonics of the initial Fourier modes on the long 
string are present and when not all the modes interact with all of the other modes. In this
case, it was shown that the minimum wavelength is given by
 \ba
 \lambda_{min}=  (G \mu_{1} )^{n} \, t_{0}
 \ea
where $n=3/2, 5/2$ for radiation and matter dominated eras, respectively.

In this work we have considered GW emission from three co-planar strings 
forming a junction and assuming that excitations are originally generated on one string
leg. It would be interesting to  generalize this exercise to more realistic cases 
of an arbitrary number of strings in a junction when incoming waves of arbitrary frequencies 
and amplitudes are excited on each string. It would be interesting to see if the results 
of \cite{OS} hold, where it was shown that GW emission from  left-moving and right-moving 
wave-trains on an infinite string is zero if the wave-trains have significantly different 
frequencies and amplitudes.

\section{Acknowledgments}

We would like to thank E. Copeland, T. Kibble, X. Siemens  and D. Steer for useful comments.
At McGill, this research has been supported by NSERC
under the Discovery Grant program. R.B. is also
supported by the Canada Research Chairs program. R.B.
wishes to thank the Theory Division of the Institute
of High Energy Physics in Beijing for their
wonderful hospitality during the final stages of this
project.


\vspace{1cm}

\large{\bf Appendix :} \,  {\bf Higher powers of $s_{i}(t)$ in $\delta T_{\mu \nu}(k)$}

\vspace{0.5cm}
In this Appendix we show that the integrals containing the powers of $s_{i}(t)$ in 
$\delta T_{\mu \nu}(k)$ in Eq. (\ref{Tmuk2s}) vanish.  To see this, consider the term 
quadratic in $s_{i}^{2}$ when the corresponding integral is
\ba
&& \int_{-\infty}^{\infty} \frac{ \ell^{2} \, d \ell }{\ell - i \varepsilon} 
  \int du \, dv \, s_{i}^{2}   e^{i \Kp v/2}  e^{i \Km u/2}  \nonumber\\
 && \sim
  \int_{-\infty}^{\infty} \frac{ \ell^{2} \, d \ell }{\ell - i \varepsilon}   \int du \, dv \,  e^{i \Kp v/2}  e^{i \Km u/2}  \sin^{2} \left(  \frac{\kappa}{2} (v-u)    \right) \, .
\ea
The integral further simplifies to
\ba
\label{bracket}
 \int_{-\infty}^{\infty} \frac{ \ell^{2} \, d \ell }{\ell - i \varepsilon}   \int du \, dv \,  e^{i \Kp v/2}  e^{i \Km u/2}  ( 1- \cos \kappa u \,  \cos \kappa v - \sin \kappa u \,  \sin \kappa v) \, .
\ea
The first term in the above bracket gives
\ba
 &&\int_{-\infty}^{\infty} \frac{ \ell^{2} \, d \ell }{\ell - i \varepsilon}   \int du \, dv \,  e^{i \Kp v/2}  e^{i \Km u/2}  \nonumber\\
 && \sim  
  \int_{-\infty}^{\infty} \frac{ \ell^{2} \, d \ell }{\ell - i \varepsilon}  \, \frac{\sin (\Kp L/4)}{\Kp} \, 
  \frac{\sin (\Km L/4)}{\Km} \, .
\ea
This integral has three poles at  $\ell=i\varepsilon$, $\Kp=0$ and $\Km=0$. However, the 
residues at all three poles vanish. Performing the integrals for the other two terms in the bracket 
in Eq. (\ref{bracket}), one can check that the residues at the poles vanish.

In conclusion, the integral in Eq. (\ref{bracket}) and correspondingly the integral in 
Eq. (\ref{Tmuk2s}) containing terms of second power in $s_{i}(t)$ vanish. Following the 
same strategy, one can check that the integral in Eq. (\ref{Tmuk2s}) containing terms 
of linear power in $s_{i}(t)$ also vanishes.



\section*{References}
\begin{thebibliography}{}

\bibitem{HenryTye:2006uv}
  S.~H.~Henry Tye,
  ``Brane inflation: String theory viewed from the cosmos,''
  Lect.\ Notes Phys.\  {\bf 737}, 949 (2008)
  [arXiv:hep-th/0610221].


\bibitem{Kibble:2004hq}
  T.~W.~B.~Kibble,
  ``Cosmic strings reborn?,'' astro-ph/0410073.
  
  \bibitem{ACD}
A.~C.~Davis and T.~W.~B.~Kibble,
  ``Fundamental cosmic strings,''
  Contemp.\ Phys.\  {\bf 46}, 313 (2005)
  [arXiv:hep-th/0505050].
 
 \bibitem{Mairi}
  M.~Sakellariadou,
  ``Cosmic Superstrings,''
  arXiv:0802.3379 [hep-th].
  
\bibitem{Shlaer:2005ry}
  B.~Shlaer and M.~Wyman,
  ``Cosmic superstring gravitational lensing phenomena: Predictions for
  networks of (p,q) strings,''
  Phys.\ Rev.\  D {\bf 72}, 123504 (2005)
  [arXiv:hep-th/0509177].
  
\bibitem{Copeland:2006eh}
  E.~J.~Copeland, T.~W.~B.~Kibble and D.~A.~Steer,
  ``Collisions of strings with Y junctions,''
  Phys.\ Rev.\ Lett.\  {\bf 97}, 021602 (2006)
  [arXiv:hep-th/0601153].

\bibitem{Copeland:2006if}
  E.~J.~Copeland, T.~W.~B.~Kibble and D.~A.~Steer,
  ``Constraints on string networks with junctions,''
  Phys.\ Rev.\  D {\bf 75}, 065024 (2007)
  [arXiv:hep-th/0611243].

\bibitem{Copeland:2007nv}
  E.~J.~Copeland, H.~Firouzjahi, T.~W.~B.~Kibble and D.~A.~Steer,
  ``On the Collision of Cosmic Superstrings,''
  Phys.\ Rev.\  D {\bf 77}, 063521 (2008)
  [arXiv:0712.0808 [hep-th]].

\bibitem{Brandenberger:2007ae}
  R.~Brandenberger, H.~Firouzjahi and J.~Karouby,
  ``Lensing and CMB Anisotropies by Cosmic Strings at a Junction,''
  Phys.\ Rev.\  D {\bf 77}, 083502 (2008)
  [arXiv:0710.1636 [hep-th]].

\bibitem{Avgoustidis:2007aa}
  A.~Avgoustidis and E.~P.~S.~Shellard,
  ``Velocity-Dependent Models for Non-Abelian/Entangled String Networks,''
  arXiv:0705.3395 [astro-ph].

\bibitem{Rajantie:2007hp}
  A.~Rajantie, M.~Sakellariadou and H.~Stoica,
  ``Numerical experiments with p F- and q D-strings: the formation of (p,q)
  bound states,''
  JCAP {\bf 0711}, 021 (2007)
  [arXiv:0706.3662].

\bibitem{Urrestilla:2007yw}
  J.~Urrestilla and A.~Vilenkin,
  ``Evolution of cosmic superstring networks: a numerical simulation,''
  JHEP {\bf 0802}, 037 (2008)
  [arXiv:0712.1146 [hep-th]].

\bibitem{Davis:2008kg}
  A.~C.~Davis, W.~Nelson, S.~Rajamanoharan and M.~Sakellariadou,
  ``Cusps on cosmic superstrings with junctions,''
  arXiv:0809.2263 [hep-th].

\bibitem{Leblond:2007tf}
  L.~Leblond and M.~Wyman,
  ``Cosmic Necklaces from String Theory,''
  Phys.\ Rev.\  D {\bf 75}, 123522 (2007)
  [arXiv:astro-ph/0701427].

\bibitem{Dasgupta:2007ds}
  K.~Dasgupta, H.~Firouzjahi and R.~Gwyn,
  ``Lumps in the throat,''
  JHEP {\bf 0704}, 093 (2007)
  [arXiv:hep-th/0702193].


\bibitem{Suyama:2008ty}
  T.~Suyama,
  ``Exact gravitational lensing by cosmic strings with junctions,''
  Phys.\ Rev.\  D {\bf 78}, 043532 (2008)
  [arXiv:0807.4355 [astro-ph]].



\bibitem{Hindmarsh:1994re}
  M.~B.~Hindmarsh and T.~W.~B.~Kibble,
  ``Cosmic strings,''
  Rept.\ Prog.\ Phys.\  {\bf 58}, 477 (1995)
  [arXiv:hep-ph/9411342].

\bibitem{book}
``Cosmic Strings and Other Topological Defects,'' A. Vilenkin and E. P. S. Shellard,
Cambridge University Press, 1994.

\bibitem{Damour:2004kw}
  T.~Damour and A.~Vilenkin,
  ``Gravitational radiation from cosmic (super)strings: Bursts, stochastic
  background, and observational windows,''
  Phys.\ Rev.\  D {\bf 71}, 063510 (2005)
  [arXiv:hep-th/0410222].

\bibitem{Xiemens} 
X.~Siemens, V.~Mandic and J.~Creighton,
  ``Gravitational wave stochastic background from cosmic (super)strings,''
  Phys.\ Rev.\ Lett.\  {\bf 98}, 111101 (2007)
  [arXiv:astro-ph/0610920];\\
  X.~Siemens, J.~Creighton, I.~Maor, S.~Ray Majumder, K.~Cannon and J.~Read,
  ``Gravitational wave bursts from cosmic (super)strings: Quantitative
  analysis and constraints,''
  Phys.\ Rev.\  D {\bf 73}, 105001 (2006)
  [arXiv:gr-qc/0603115].
  
\bibitem{Sakellariadou:1990ne}
  M.~Sakellariadou,
  ``Gravitational waves emitted from infinite strings,''
  Phys.\ Rev.\  D {\bf 42}, 354 (1990)
  [Erratum-ibid.\  D {\bf 43}, 4150 (1991)].

\bibitem{Hindmarsh:1990xi}
  M.~Hindmarsh,
  ``Gravitational radiation from kinky infinite strings,''
  Phys.\ Lett.\  B {\bf 251}, 28 (1990).

\bibitem{OS}
  X.~Siemens and K.~D.~Olum,
  ``Gravitational radiation and the small-scale structure of cosmic  strings,''
  Nucl.\ Phys.\  B {\bf 611}, 125 (2001)
  [Erratum-ibid.\  B {\bf 645}, 367 (2002)]
  [arXiv:gr-qc/0104085].
 
\bibitem{Weinberg}
S. Weinberg, ``{Gravitation and cosmology: principles and applications of the general theory of relativity}'', Wiley, New York, 1972.

\bibitem{Siemens:2002dj}
  X.~Siemens, K.~D.~Olum and A.~Vilenkin,
  ``On the size of the smallest scales in cosmic string networks,''
  Phys.\ Rev.\  D {\bf 66}, 043501 (2002)
  [arXiv:gr-qc/0203006].

\bibitem{Polchinski:2007rg}
  J.~Polchinski and J.~V.~Rocha,
  ``Cosmic string structure at the gravitational radiation scale,''
  Phys.\ Rev.\  D {\bf 75}, 123503 (2007)
  [arXiv:gr-qc/0702055].





\end{thebibliography}
\end{document}